\documentclass[aps,pre,reprint,superscriptaddress]{revtex4-2}
\usepackage{amsmath,graphicx,siunitx,xcolor,hyperref,physics2}
\usephysicsmodule{ab.legacy,braket,ab}
\hypersetup{
colorlinks,
linkcolor={blue!50!black},
citecolor={blue!50!black},
urlcolor={blue!80!black}}
\newcommand{\Nup}{N_\text{up}}

\begin{document}
\title{Particle pinning as a method to manipulate marginal stability}
\author{Kumpei Shiraishi}
\email{kumpei.shiraishi@umontpellier.fr}
\affiliation{Laboratoire Charles Coulomb (L2C), Universit\'e de Montpellier, CNRS, 34095 Montpellier, France}
\author{Yusuke Hara}
\affiliation{Graduate School of Arts and Sciences, University of Tokyo, Komaba, Tokyo 153-8902, Japan}
\date{\today}
\begin{abstract}
We study the critical behavior of low-frequency vibrations of packings with pinned particles near the jamming point.
Soft modes form a plateau in the density of states and its frequency is controlled by the contact number as the ordinary jamming transition.
The spatial structure of these modes is not largely affected by pins.
Below the plateau, the non-Debye scaling predicted by mean-field theories and quasi-localized modes breaks down depending on the pinning procedures.
We comprehensively explain these behaviors by the impact of pinning operations on the marginal stability of the packings.
\end{abstract}
\maketitle

\section{Introduction}
The jamming transition, wherein constituent particles solidify into a disordered structure, has garnered significant attention due to its relevance in various soft matter materials.
Extensive investigations have focused on a system composed of frictionless spherical particles, which have provided valuable insights into the nature and mechanisms underlying this phenomenon~\cite{O_Hern_2003,van_Hecke_2010,Liu_2010}.
A key concept is the isostaticity, where the number of constraints matches the number of degrees of freedom within the packing.
This condition is satisfied at the transition~\cite{O_Hern_2003} where the pressure $p$ vanishes.
The excess contact number $\delta z = z - z_\text{iso}$, which measures the deviation from isostaticity, follows a scaling law of $\delta z \sim p^{1/2}$~\cite{O_Hern_2003}.
Importantly, many packing properties exhibit scaling behaviors governed by $\delta z$~\cite{Wyart_EPL_2005,Wyart_PRE_2005,Wyart_EPL_2010,DeGiuli_2014,Yan_2016}.

\begin{figure}
\centering
\includegraphics[width=\linewidth]{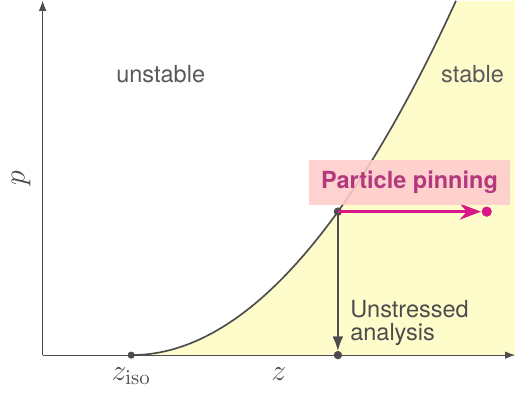}
\caption{Stability phase diagram of jammed harmonic spheres and the summary of this study.
The diagram is divided into stable and unstable phases by the marginal stability line $\delta z=Cp^{1/2}$.
Ordinary jammed packings (the circle at the center of the figure) are located on this line and thus they are marginally stable.
In previous studies, unstressed analysis has been used to stabilize the packing by moving it away from this line.
In this study, we show that particle pinning can stabilize the packing in another direction from unstressed analysis and that the distance from the marginal stability line can be manipulatively controlled.}
\label{fig:phase_diagram}
\end{figure}

Recent advances in mean-field theories~\cite{DeGiuli_2014,Franz_2015} and simulations of this system~\cite{Mizuno_2017} have provided insights into the low-frequency region of the vibrational density of states (VDOS), denoted as $g(\omega)$, unveiling its functional form as follows~\cite{Mizuno_2017}:
\begin{align}
g(\omega) =
\begin{cases}
\alpha_* \omega^0                                        & \omega \gg \omega^*, \\
\alpha_\text{BP} (\omega/\omega^*)^2                     & \omega^* \gg \omega \gg \omega_0, \\
A_0 \omega^{d-1} + \alpha_\text{loc} (\omega/\omega^*)^4 & \omega_0 \gg \omega.
\end{cases}
\end{align}
At higher frequencies $\omega \gg \omega^*$, the VDOS presents a sustained plateau up to an onset frequency $\omega^*$, which follows $\omega^* \sim \delta z$~\cite{Silbert_2005,Wyart_PRE_2005}.
Extensive study reveals the origin of this scaling in the isostaticity of packings~\cite{Wyart_EPL_2005,Wyart_PRE_2005,Wyart_EPL_2010,DeGiuli_2014,Yan_2016}.
Interestingly, this relation holds for dimer-shaped nonspherical particles~\cite{Shiraishi_2019,Shiraishi_2020}.
Plateau modes extend across the system but display a disordered pattern distinct from the Debye-predicted phonon modes, known as anomalous modes~\cite{Silbert_Liu_Nagel_2009}.
Below $\omega^*$, the VDOS follows the non-Debye scaling $g(\omega) \propto \omega^2$ in contrast to the Debye prediction $g(\omega) \propto \omega^{d-1}$.
Mean-field theories predict~\cite{DeGiuli_2014,Franz_2015} and simulations confirm this scaling for both three~\cite{Mizuno_2017} and higher dimensions~\cite{Charbonneau_2016,Shimada_2020}.
Furthermore, below a specific frequency $\omega_0$, quasi-localized modes emerge, with $g(\omega)$ contributing proportionally to $\omega^4$ in finite dimensions~\cite{Mizuno_2017}, in addition to the Debye contribution $\omega^{d-1}$ predicted by mean-field theories~\cite{DeGiuli_2014,Franz_2015}.
The origin of these properties below $\omega^*$ is attributed to another facet of packings termed marginal stability~\cite{Wyart_PRE_2005,DeGiuli_2014,Franz_2015,Mizuno_2017}.
This trait implies that packings are poised on the verge of instability.
Marginal stability and isostaticity are usually grasped on a phase diagram spanned by $z$ and $p$ axes~\cite{Wyart_PRE_2005,DeGiuli_2014}, which is illustrated in Fig.~\ref{fig:phase_diagram}.
Packings are situated on the marginal stability line $\delta z=Cp^{1/2}$ that divides stable and unstable phases ($C$ is a constant).
The intersection of this line and $z$-axis ($p=0$) marks the isostaticity point.

Quite recently, considerable attention has been directed towards the interplay between the degrees of freedom and stability of glass configurations~\cite{Brito_2018,Kapteijns_2019,Hagh2022,Ozawa2023}, stimulated by advancements in both numerical~\cite{BerthierReichman2023} and experimental~\cite{Ediger_2017} methodologies, namely the swap Monte Carlo and vapor deposition techniques, respectively.
Both methods achieve unprecedented stability by adding or freezing degrees of freedom;
in the swap Monte Carlo, non-translational degrees of freedom are explicitly taken into account by allowing changes in particle diameters, and in the vapor deposition techniques, molecules at the free surface have more movable degrees of freedom than molecules in the bulk~\cite{Hagh2022,Ozawa2023}.
And, vibrations in the low-frequency region below $\omega^*$ bear a direct relationship to the stability~\cite{Wang_low_freq_2019}.
Within this context, jammed particles, where the constraints imposed on the degrees of freedom precisely control the characteristic frequency $\omega^*$ of low-frequency vibrations, provide an ideal platform to advance research.
An interesting approach is the random pinning method~\cite{Brito_2013}.
This method has been employed to explore the slow dynamics of supercooled liquids~\cite{Kim_2003,Kim2009} and the glass transition~\cite{Cammarota_2012,Kob_2013,Ozawa_2015}, and recently, it has been shown to have a potential to probe low-frequency vibrations of glasses~\cite{Angelani_2018,Shiraishi2022Ideal,Shiraishi2023pin2D}.
While the method has been utilized to study structural properties of jammed packings~\cite{PhysRevE.86.061301,Graves2016,Zhang2022pins,Zhang2022Revealing}, its application to their vibrational properties remains limited.
Brito \textit{et al.}~applied this method to the hard-sphere system below jamming and demonstrated the generation of hyperstatic packings depending on the pinning protocol~\cite{Brito_2013}.
However, their study focused solely on vibrations near $\omega^*$.
Given the recent progress in understanding the frequency region lower than $\omega^*$, it is imperative to investigate how pinning and the consequent changes in isostaticity and marginal stability affect vibrations in this relevant regime.

Here, we investigate the impact of pinned particles on low-frequency vibrations by introducing two methods for generating packings.
Our findings reveal the preservation of behaviors governed by isostaticity at $\omega > \omega^*$, irrespective of the pinning method employed.
Conversely, at $\omega < \omega^*$, the pinning method emerges as a factor that could move the system away from marginal stability, consequently influencing the associated vibrational properties.
Through our investigation, we provide compelling evidence that the concept of isostaticity and marginal stability offers a unified explanation for the effects of pinning on the properties of jammed packings.

\section{Methods}
We consider random packings of $N$ frictionless particles with identical mass $m$ enclosed in a square box with periodic boundary conditions in two (2D) and three (3D) dimensions.
Particles interact with a repulsive harmonic potential
\begin{align}
v(r_{ij}) = \frac{\epsilon}{2} \pab{1 - \frac{r_{ij}}{\sigma_{ij}}}^2
\end{align}
only if the distance $r_{ij}$ between particles $i$ and $j$ is smaller than the sum $\sigma_{ij}$ of their radii $\sigma_i/2, \sigma_j/2$.
We use 50:50 binary mixtures of particles whose diameters are $\sigma$ and $1.4\sigma$.
Mass, lengths, and energies are measured in units of $m$, $\sigma$, and $\epsilon$, respectively.

We employ two protocols to generate jammed packings with pinned particles.
Here, we denote the fraction of pinned particles as $c$ ($0 \leq c \leq 1$) and the number of unpinned particles as $\Nup=(1-c)N$.
For the Compression-Pinning (CP) protocol, packings with a given target pressure $p$ are initially generated through the global compression/decompression~\cite{GoodrichThesis}, then pinned particles are selected in these packings.
Meanwhile, in the Pinning-Compression (PC) protocol, pinned particles are selected at a stage of mechanically stable configurations at a high enough packing fraction $\varphi_\text{init}=1.0$.
Then, given pressures are achieved in the same way as the CP protocol, except that pinned particles are moved only affinely and fixed during relaxations to mechanical equilibrium.
To select pinned particles, we used template configurations at $p = \num{e-1}$ with each number of particles $cN$.
The advantage of this method is that we can identify pinned particles in a uniform manner and avoid yielding regions where pinned particles are too dense.
This procedure is detailed in Refs.~\cite{Kob_2013,Ozawa_2015,Shiraishi2022Ideal}.
For the energy minimization, the FIRE algorithm~\cite{Guenole_2020} is used (the termination condition of the algorithm is $\max_i F_i < \num{e-12}$).
After packings are generated, we recursively remove the rattler particles whose number of contacts is smaller than the spatial dimension $d$.
The ensemble sizes of each protocol are 2000 for the CP protocol ($N=1000$), 1000 for the CP protocol ($N=4000$), and 1000 for the PC protocol ($N=1000$) in 2D.

Finally, we performed the vibrational analysis with pinned particles~\cite{Shiraishi2022Ideal}.
The vibrational analysis is performed for $\Nup$ unpinned particles with the presence of pinned particles; the formulation of the analysis can be found in Ref.~\cite{Shiraishi2022Ideal}.
In this analysis, the dynamical matrix $\mathcal{M}$ (size $d\Nup \times d\Nup$) is calculated whose off-diagonal elements are given as
\begin{align}
\mathcal{M}_{ij} = - v^{\prime\prime}(r_{ij}) \hat{\boldsymbol{r}}_{ij} \otimes \hat{\boldsymbol{r}}_{ij} - \frac{v^\prime(r_{ij})}{r_{ij}} \pab{\mathcal{I} - \hat{\boldsymbol{r}}_{ij} \otimes \hat{\boldsymbol{r}}_{ij}}, \label{dynmat}
\end{align}
and diagonal elements are given as
\begin{align}
\mathcal{M}_{ii} = - \sum_{j \in \partial i} \mathcal{M}_{ij}.
\end{align}
Here, $\hat{\boldsymbol{r}}_{ij}$ is a unit vector along $\boldsymbol{r}_i - \boldsymbol{r}_j$, $\partial_i$ represents the set of neighboring particles of $i$, and $\mathcal{I}$ is the identity matrix whose size is $d \times d$.
The diagonalization problem of the dynamical matrix is solved numerically using the Eigen package~\cite{eigenweb}.
The eigenvalues and eigenvectors are denoted as $\lambda_k$ and $\boldsymbol{e}_k = \pab{\boldsymbol{e}_k^1, \dots, \boldsymbol{e}_k^{\Nup}}$ respectively for each eigenmode $k = 1, 2, \dots, d\Nup$.
The eigenfrequency $\omega_k$ is determined as $\omega_k = \sqrt{\lambda_k}$.

\section{Contact number}
\begin{figure}
\centering
\includegraphics[width=\linewidth]{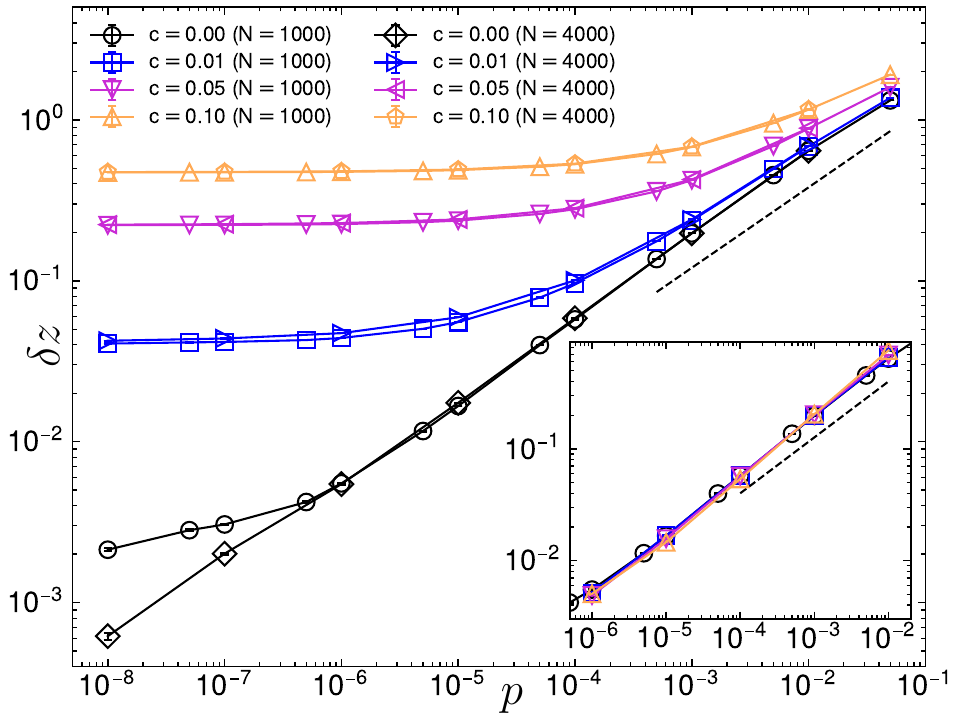}
\caption{The excess contact number versus pressure in 2D packings generated by the CP protocol.
Inset: $\delta z$ vs $p$ for packings with $N=1000$ generated by the PC protocol.
The legends are the same as the main panel.
The error bars are estimated by the bootstrap method and the dashed lines represent $\delta z \sim p^{1/2}$.}
\label{fig:dz}
\end{figure}

First, we consider the definition of the isostatic condition with pinned particles.
In the normal jammed packings in $d$ dimensions, the isostatic condition is written as~\cite{Goodrich_2012}
\begin{align}
\frac{zN}{2} = dN-d
\end{align}
for finite systems.
The left-hand side is the number of constraints in the packing and $z$ is the average contact number.
The right-hand side is the number of degrees of freedom.
The subtraction of $d$ comes from $d$ zero-frequency modes.
The excess contact number $\delta z = z - z^N_\text{iso}$ to the isostatic number $z^N_\text{iso} = 2d(1-1/N)$ shows the scaling $\delta z \sim p^{1/2}$ near the jamming transition (Fig.~\ref{fig:dz}), which has been widely observed in past studies~\cite{O_Hern_2003,Goodrich_2012}.

When pinned particles are introduced, the isostatic condition is altered.
Because the pinned particles break the translational invariance, pinned systems no longer show $d$ zero modes~\cite{Angelani_2018,Shiraishi2022Ideal,Shiraishi2023pin2D}.
Therefore, the isostatic condition with pinning becomes
\begin{align}
\frac{z\Nup}{2} = d\Nup,\label{PinnedMaxwell}
\end{align}
and the excess contact number is defined as $\delta z = z - 2d$.

Unlike the unpinned system, the pinned system comprises two distinct particle types: pinned particles and unpinned (movable) particles.
Hence, when calculating the excess contact number $\delta z$ of the pinned system based on the above definition, consideration is needed whether particles consisting the given contact are pinned or not.
In the ordinary contact number analysis, a single contact between two particles constrains the relative motion of both particles.
However, in the pinned system, a contact constrains only the motion of the movable particle if one of the particles is pinned.
Therefore, contacts formed between the movable and pinned particles must be considered as imposing twice as much constraint as contacts between two movable particles.
As a result, the contact number of movable particle $i$ in the pinned system is defined as
\begin{align}
z_i = z_{i, \text{up}} + 2z_{i, \text{p}},
\end{align}
where $z_{i, \text{up}}$ is the number of contacts between unpinned particles and $z_{i, \text{p}}$ is that of between pinned particles.

Based on this discussion, we calculated the excess contact number of 2D packings (Fig.~\ref{fig:dz}).
In contrast to unpinned packings that exhibit the scaling $\delta z \sim p^{1/2}$ and hence $z$ approach the isostatic value as $p \to 0$, $z$ of pinned packings generated using the CP protocol do not converge to $z_\text{iso} = 2d$.
Instead, they possess a greater number of contacts near jamming, indicating a hyperstatic nature at the transition.
This behavior aligns with observations made in hard-sphere systems below jamming~\cite{Brito_2013}, and can easily be understood that the CP protocol forcibly removes degrees of freedom from packings that are made to satisfy the isostatic condition.
For the CP packings, the total number of contacts approaches $dN$ as $p \to 0$ instead of $d\Nup$ because the packings are originally generated to satisfy the isostatic condition of the unpinned system.
However, pinning operations of the CP protocol reduces the number of degrees of freedom while keeping the number of constraints, causing a mismatch in Eq.~\eqref{PinnedMaxwell}.
Therefore, the average contact number $z$ approaches to $2d\pab{N/\Nup} = 2d/\pab{1-c}$.
From this discussion, we can evaluate the excess contact number of CP packings at $p \to 0$ as $\delta z = 2d/\pab{1-c}-2d \approx 2dc$.
Additionally, we observe that the presence of pinned particles erases the finite-size effect at $p \leq \num{e-7}$ that is due to a condition of packing has positive bulk modulus~\cite{Goodrich_2012} because it maintains $\delta z$ at large values.
In contrast, PC packings do not exhibit the hyperstatic behavior (inset of Fig.~\ref{fig:dz}).
The scaling $\delta z \sim p^{1/2}$ holds across the entire pressure range $\left[\num{e-6}, \num{e-2}\right]$ that we have explored, indicating that PC packings are isostatic at the transition.
The consistency of these findings is confirmed in 3D packings.

It is worth noting that a subset of PC packings with $p \leq \num{e-5}$ and $c=0.10$ contains rattlers, where almost all unpinned particles are involved.
The occurrence rate of such packings was found to be \SI{3.8}{\percent} at $p=\num{e-6}$ and $c=0.10$ in 2D.
We exclude them from our analysis.
The properties of these loosely packed configurations will be investigated in future studies.

\section{Plateau in the density of states}
\begin{figure}
\centering
\includegraphics[width=\linewidth]{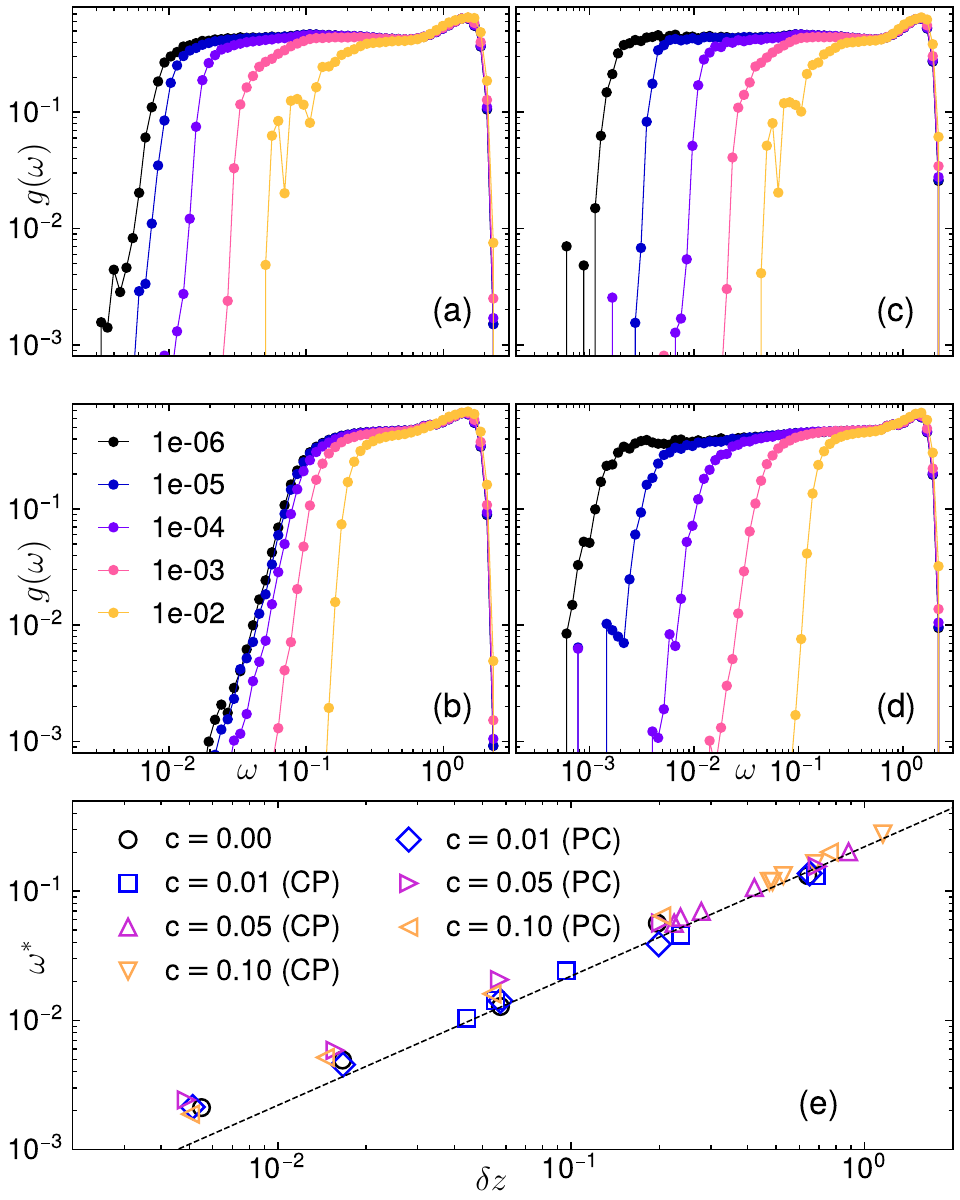}
\caption{The vibrational density of states of $p=\numrange{e-6}{e-2}$ with
(a) $c=0.01$ by CP protocol,
(b) $c=0.10$ by CP protocol,
(c) $c=0.01$ by PC protocol, and
(d) $c=0.10$ by PC protocol.
2D packings with $N=1000$ are used for this analysis.
(e) The characteristic frequency $\omega^*$ of the onset of the plateau.
The dashed line represents $\omega^* \sim \delta z$.}
\label{fig:vib}
\end{figure}

Next, we analyze the vibrational properties of packings with pinning.
In the following, we perform vibrational analysis on the unstressed system~\cite{Wyart_PRE_2005}, an analysis in which only the stiffness terms of the interaction are retained and the repulsive forces, i.e., the second term (pre-stressed terms) in Eq.~\eqref{dynmat} are dropped when calculating the dynamical matrix $\mathcal{M}$.
The critical behavior of the plateau of the VDOS is not affected by the forces which are vanishingly small near the transition~\cite{Wyart_PRE_2005}.
The forces make the system mechanically unstable and only affect the modes below the plateau~\cite{Wyart_PRE_2005,Lerner_2014,Mizuno_2017,Shimada_Mizuno_Wyart_Ikeda_2018}.
Here, we focus on the plateau behavior and employ the unstressed system.

We consider the VDOS, which is calculated as
\begin{align}
g(\omega) = \frac{1}{N_\text{mode}} \sum_k \delta(\omega - \omega_k)
\end{align}
where $N_\text{mode}$ is the number of all eigenmodes and $\delta(x)$ is the Dirac delta function.
For CP packings with $c=0.01$, the VDOS exhibits a distinct plateau at low frequencies for all pressures investigated (Fig.~\ref{fig:vib} (a)).
At the highest pressure, $p=\num{e-2}$, discernible peaks attributed to phonons can be observed below the plateau.
As the system approaches jamming $p \to 0$, the plateau extends all the way to zero frequency.
Conversely, for $c=0.10$, the situation differs (Fig.~\ref{fig:vib} (b)).
While $g(\omega)$ still displays a plateau at low frequencies, its extension to zero frequency as $p \to 0$ is limited.
On the other hand, $g(\omega)$ of PC packings exhibits a plateau that similarly extends to zero frequency for both $c=0.01$ (Fig.~\ref{fig:vib} (c)) and $c=0.10$ (Fig.~\ref{fig:vib} (d)).
The only noticeable distinction is the absence of phonon peaks below the plateau when $c=0.10$ and $p=\num{e-2}$, which is also observed in CP packings (Figs.~\ref{fig:vib} (a), (b)).
The disappearance of phonons at large $c$ has been well studied~\cite{Angelani_2018,Shiraishi2022Ideal,Shiraishi2023pin2D}.

The plateau is characterized by the onset frequency $\omega^*$, which is known to be controlled by $\delta z$~\cite{Wyart_PRE_2005}.
In our investigation, we have also computed this relationship.
It has been consistently observed that $\omega^*$ scales with $\delta z$ in all cases examined (Fig.~\ref{fig:vib} (e)).
Figure~\ref{fig:dz} illustrates that the CP packings lack the isostaticity at jamming.
This behavior is also reflected in $g(\omega)$: despite a decrease in pressure, CP packings remain distant from the isostaticity in terms of $\delta z$.
Consequently, their low-frequency plateau does not extend to zero frequency (Fig.~\ref{fig:vib} (b)).
Nonetheless, if we consider the plateau as a function of $\delta z$, representing the distance from $z_\text{iso}$, the behavior of $\omega^*$ aligns with the well-established scaling $\omega^* \sim \delta z$~\cite{Wyart_PRE_2005,Silbert_2005}, even with the presence of pinned particles.

\section{Spatial structure of anomalous modes}
\subsection{Visualization of anomalous modes}
\begin{figure}
\centering
\includegraphics[width=\linewidth]{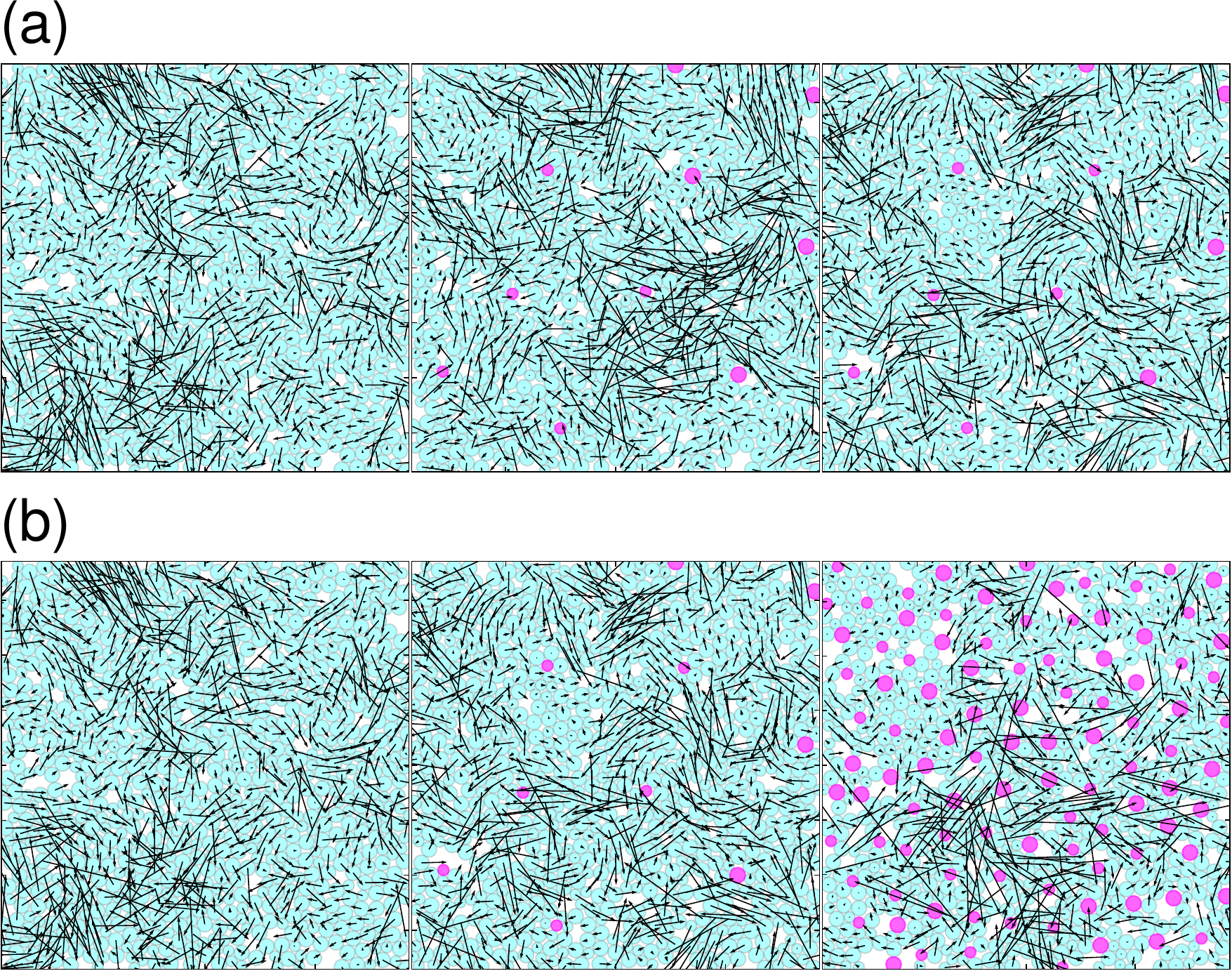}
\caption{The visualization of plateau modes.
(a) Plateau modes in unpinned ($c=0.00$), pinned ($c=0.01$, CP), and pinned ($c=0.01$, PC) packings are shown (from left to right).
(b) Plateau modes at $c=0.00, 0.01, 0.10$ with PC protocol are shown (from left to right).
Packings ($N=1000$) at $p=\num{e-4}$ are used for both panels and frequencies of each mode are $\omega=\num{4.5e-2}$ close to $\omega^*$.
The particles in magenta represent pinned particles.}
\label{fig:visualize}
\end{figure}

We further investigate the spatial structure of plateau modes (anomalous modes) near $\omega^*$.
Figure \ref{fig:visualize} (a) illustrates the eigenvector fields of the modes of unpinned packing and pinned packings at $c=0.01$ of both protocols.
In the absence of pinning, anomalous modes are extended throughout the system, displaying strong disorder~\cite{Silbert_Liu_Nagel_2009}.
This disordered spatial feature is distinct from plane waves.
With the introduction of pinned particles, we observe that the mode preserves its extended and disordered character.
This observation remains consistent whether employing the CP or PC protocol.

Next, we increase the pinning fraction $c$ while limiting the focus to the PC protocol.
Here, we show the same visualization of anomalous modes of PC packings for $c=0.00, 0.01, 0.10$ in Fig.~\ref{fig:visualize} (b).
With an increasing fraction of pinned particles ($c=0.10$), the mode becomes slightly localized to a specific region in the packing.
However, unlike lowest-frequency modes in pinned Lennard-Jones systems that show localization even at the level of individual particles~\cite{Shiraishi2022Ideal,Shiraishi2023pin2D}, the visualizations in Fig.~\ref{fig:visualize} (b) do not exhibit the same level of localization.
In the next subsection, we will further analyze the spatial structure of these modes using quantitative indicators.

\subsection{Spatial correlation and participation ratio of anomalous modes}
\begin{figure}
\centering
\includegraphics[width=\linewidth]{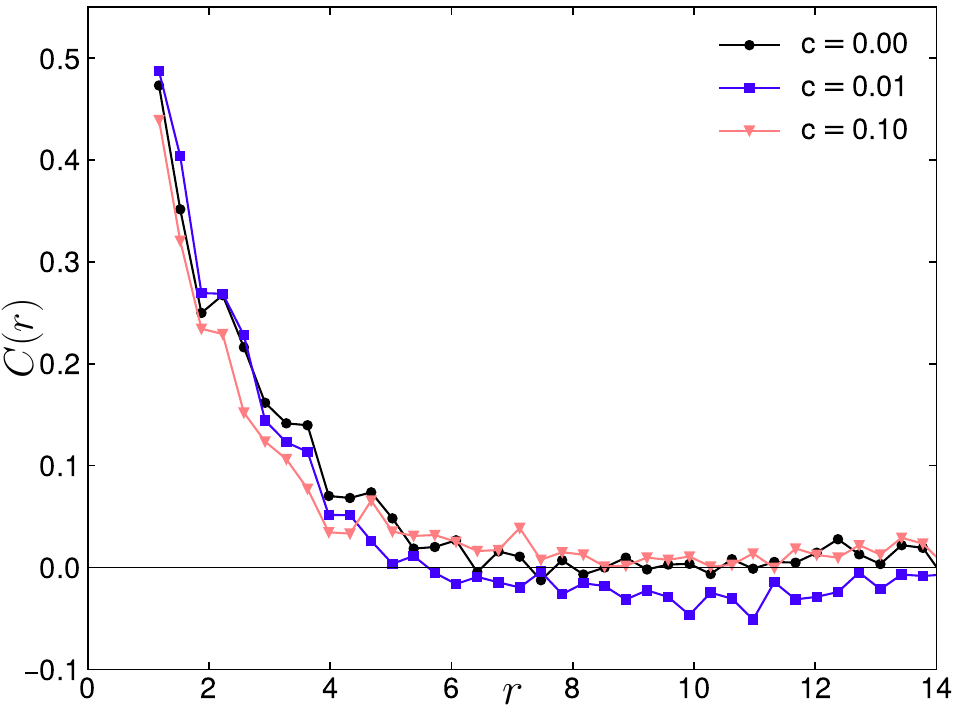}
\caption{Spatial mode correlation $C(r)$ of the plateau modes illustrated in Fig.~\ref{fig:visualize} (b).
Packings at $p=\num{e-4}$ generated through the PC protocol is used in this figure.}
\label{fig:Cr}
\end{figure}

To directly demonstrate that the spatial structure of plateau modes does not change much as the pinning fraction increases, we calculate the spatial correlation $C(r)$ for each mode~\cite{Silbert_Liu_Nagel_2009}.
$C(r)$ is defined as
\begin{align}
C(r) = \braket[1]{\hat{\boldsymbol{e}}_k^i\pab{\boldsymbol{r}_i} \cdot \hat{\boldsymbol{e}}_k^j\pab{\boldsymbol{r}_j}}_{ij},
\end{align}
where $\hat{\boldsymbol{e}}_k^i(\boldsymbol{r}_i)$ is the normalized eigenvector on particle $i$ of mode $k$ and the average is taken over the pairs whose the distance $r_{ij}$ is inside bins.
$C(r)$ offers a spatial extent of correlation in each mode $k$.
In Fig.~\ref{fig:Cr}, $C(r)$ of the modes visualized in Fig.~\ref{fig:visualize} (b) are shown.
At $c=0.00$, the correlation $C(r)$ decays quickly to zero and this is a characteristic behavior of anomalous plateau modes in jammed packings~\cite{Silbert_Liu_Nagel_2009}.
As $c$ increases, $C(r)$ does not change their decay behaviors; anomalous modes of the pinned system are also disordered, as are the standard anomalous modes of the unpinned system.
From the above observations, we conclude that the changes in spatial structures of the plateau modes undergo due to pinning are relatively slight.

To characterize spatial localization properties of plateau mode in 2D packings, we calculate the participation ratio
\begin{align}
p_k = \frac{1}{\Nup\sum_{i=1}^{\Nup}\abs{\boldsymbol{e}_k^i}^4},
\end{align}
where $\boldsymbol{e}_k^i$ is the displacement vector of particle $i$ in mode $k$.
This quantity measures the fraction of particles that participate in mode $k$~\cite{Schober_1991}.
As in the extreme cases, $p_k = 1$ ($\Nup p_k = \Nup$) for an ideal mode in which all the unpinned particles vibrate equally, and $p_k = 1/\Nup \ll 1$ ($\Nup p_k = 1$) for an ideal mode involving only one particle.

\begin{figure}
\centering
\includegraphics[width=\linewidth]{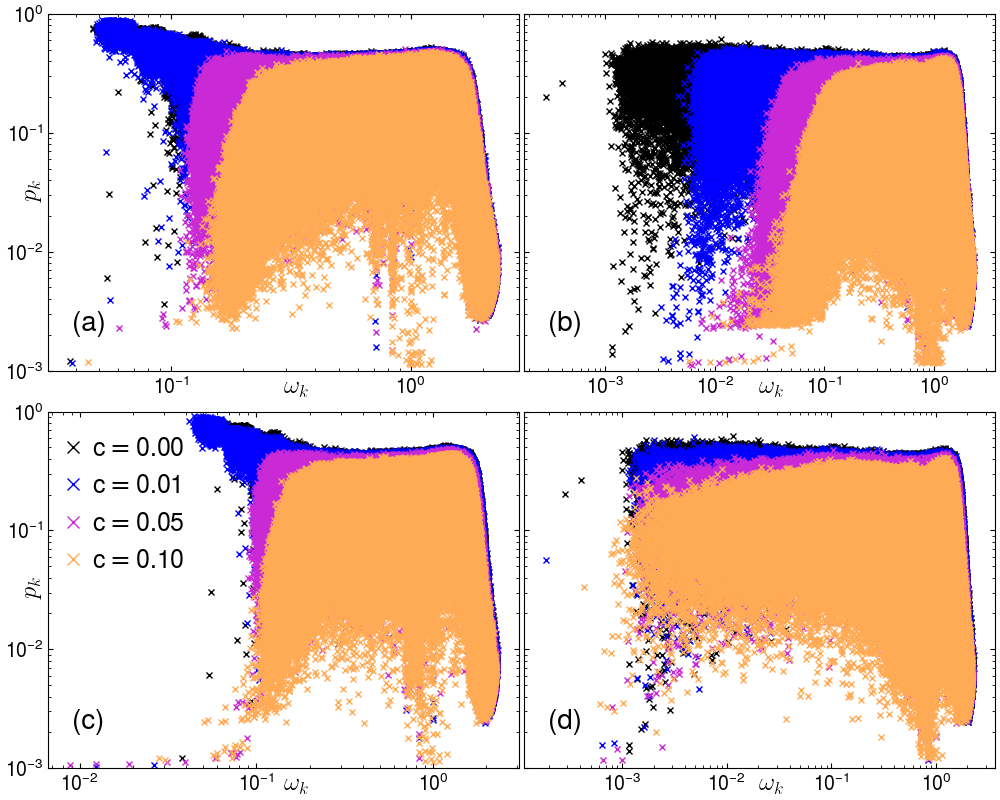}
\caption{Participation ratio of two-dimensional packings generated by the CP protocol at (a) $p=\num{e-2}$ and (b) $p=\num{e-6}$, and the PC protocol at (c) $p=\num{e-2}$ and (d) $p=\num{e-6}$.}
\label{fig:2DPR}
\end{figure}

Figure~\ref{fig:2DPR} shows the scatter plots of $p_k$ of 2D packings generated through the CP and PC protocols at $p = \num{e-2}, \num{e-6}$.
In each protocol at $p=\num{e-2}$ (Figs.~\ref{fig:2DPR} (a) and (c)), phonon modes with high $p_k$ at low frequencies disappear as $c$ increases.
This behavior is consistent with studies with the Lennard-Jones systems~\cite{Shiraishi2022Ideal,Shiraishi2023pin2D}.
At low pressure $p=\num{e-6}$ (Figs.~\ref{fig:2DPR} (b) and (d)), the plateau extends to low enough frequencies following $\omega^* \sim \delta z \sim p^{1/2}$, and therefore, phonons are not clearly visible at $c=0.00$ due to the finite-size effect~\cite{Mizuno_2017}.
However, in plateau regions, the decrease of $p_k$ of each mode is limited.
For example, in Fig.~\ref{fig:2DPR} (c), eigenmodes of $\omega_k \approx \num{2e-1}$, that is close to $\omega^*$, decrease $p_k$ from \num{4e-1} to \num{3e-1} as $c$ increases from 0.00 to 0.10.
The approximate number of particles participating in the plateau modes does not change significantly and the localization induced by the pinning is limited.
This contrasts with the low-frequency range of the Lennard-Jones systems, where their modes localize to about $p_k = 1/\Nup$, i.e., a single-particle vibrations~\cite{Shiraishi2022Ideal,Shiraishi2023pin2D}.
Note that in Fig.~\ref{fig:2DPR}, we can also observe discretized modes appeared at $\omega_k \approx 10^0$ when $c=0.10$.
We speculate that these modes are optical phonons.
Similar discretized modes at higher frequencies are also observed in dimer packings especially when the particle asphericity is weak~\cite{Shiraishi_2019,Shiraishi_2020}.
In summary, though the three analyses of eigenmode visualization, the spatial correlation $C(r)$, and the participation ratio $p_k$, we revealed that the anomalous modes that form the plateau of the VDOS do not change their spatial structure significantly even when pinned particles are introduced.

\section{Non-Debye scaling}
We shift our focus to frequencies lower than $\omega^*$, where the VDOS showcases the non-Debye scaling~\cite{Charbonneau_2016} and $\omega^4$ scaling due to quasi-localized modes~\cite{Mizuno_2017}.
For this investigation, we exclusively use data from 3D packings, as the abundance of phonons in 2D obscures the observation of non-phononic modes in this regime~\cite{Mizuno_2017}.
As mentioned earlier, the neglected forces in the unstressed analysis contribute to the emergence of non-phononic modes below $\omega^*$~\cite{Wyart_PRE_2005,Lerner_2014,Mizuno_2017,Shimada_Mizuno_Wyart_Ikeda_2018}.
Consequently, in this analysis, we consider the original stressed system to examine these modes in detail.

\begin{figure}
\centering
\includegraphics[width=\linewidth]{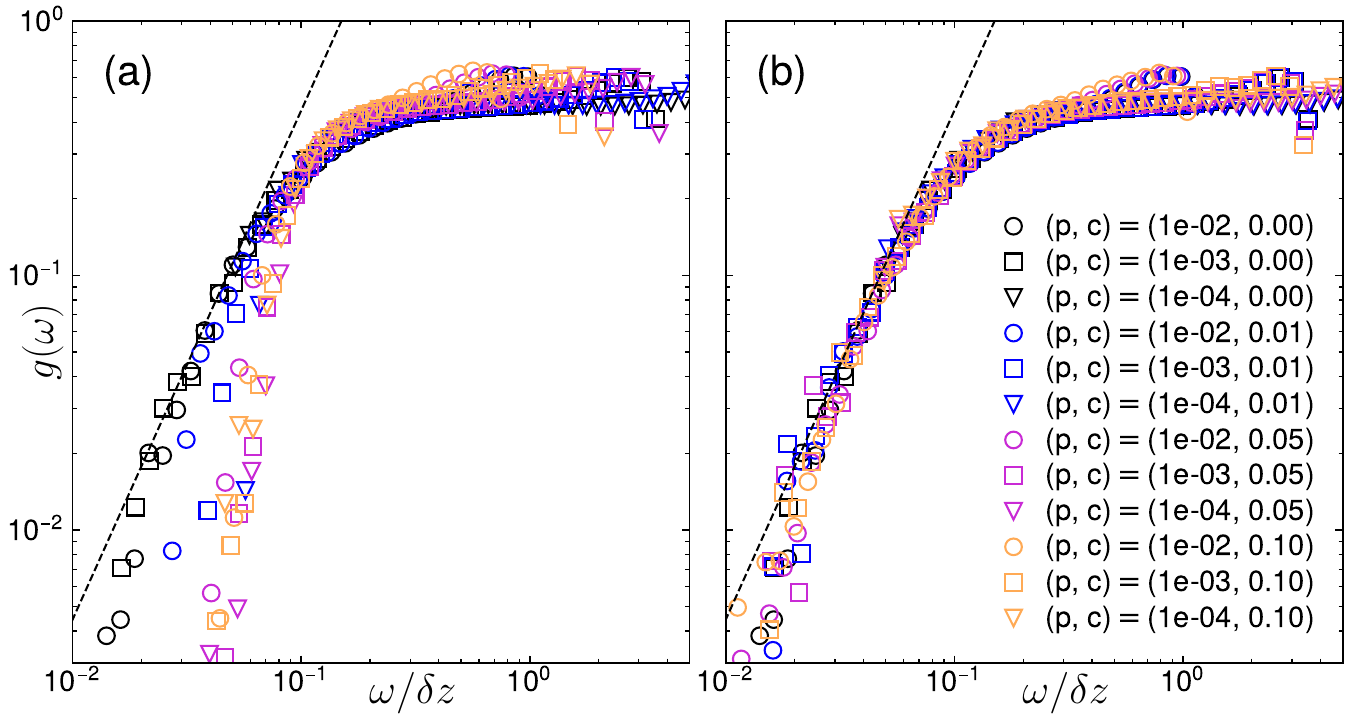}
\caption{The vibrational density of states with rescaled $\omega$.
The data is calculated from 3D packings with $N=4000$ generated by (a) CP protocol and (b) PC protocol.
The dashed lines represent $g(\omega) \propto \omega^2$.}
\label{fig:non-debye}
\end{figure}

Figure~\ref{fig:non-debye} presents the VDOS with rescaled $\omega$ for packings obtained from the CP (Fig.~\ref{fig:non-debye} (a)) and PC protocol (Fig.~\ref{fig:non-debye} (b)).
The data from the unpinned system, $c=0.00$, collapses onto a single curve through the relation $\delta z \sim \omega^*$, and exhibits the non-Debye scaling $g(\omega) \propto \omega^2$ below the plateau~\cite{Charbonneau_2016}.
However, the introduction of pinned particles disrupts the non-Debye scaling in the case of CP protocol (Fig.~\ref{fig:non-debye} (a)).
As the fraction $c$ increases, the discrepancy between the data of $c=0.00$ becomes more pronounced, and the VDOS shows a sharper decline below the plateau.
In contrast, the packings from the PC protocol maintain the non-Debye scaling even when particles are pinned (Fig.~\ref{fig:non-debye} (b)).
All the data within the range examined here converge to a master curve of the VDOS.

The breakdown of the non-Debye scaling observed in Fig.~\ref{fig:non-debye} is in line with the marginal stability of packings, a property predicted as its origin by mean-field theories~\cite{DeGiuli_2014,Franz_2015}.
In the vicinity of the jamming transition, the contact number of packings differs depending on the pinning protocol (Fig.~\ref{fig:dz}).
For the CP packings, the introduction of pinning leads to stable packings far from marginality, as the degrees of freedom are artificially reduced from the marginally stable original packings.
Consequently, the non-Debye scaling associated with marginal stability no longer holds.
An intriguing aspect is that modes contributing to both the non-Debye scaling and the plateau share spatially extended characters~\cite{Mizuno_2017}, yet respond differently to changes in marginal stability.
While the former breaks down (Fig.~\ref{fig:non-debye} (a)), the latter retains its property (Fig.~\ref{fig:vib} (e)).
This observation marks the distinct impact of marginal stability on regions below and above $\omega^*$.
Also, the data in Fig.~\ref{fig:dz} indicates an increasing degree of hyperstaticity at larger $c$, which aligns with the increasing deviation from the non-Debye behavior in Fig.~\ref{fig:non-debye} (a).
In contrast, the PC packings remain marginally stable, as evidenced by the scaling relationship $\delta z \sim p^{1/2}$, serving as the instability (marginal stability) line in the phase diagram (Fig.~\ref{fig:phase_diagram}), that holds.
Therefore, even when pinned particles are introduced, the non-Debye scaling is preserved in these packings (Fig.~\ref{fig:non-debye} (b)).

\begin{figure}
\centering
\includegraphics[width=\linewidth]{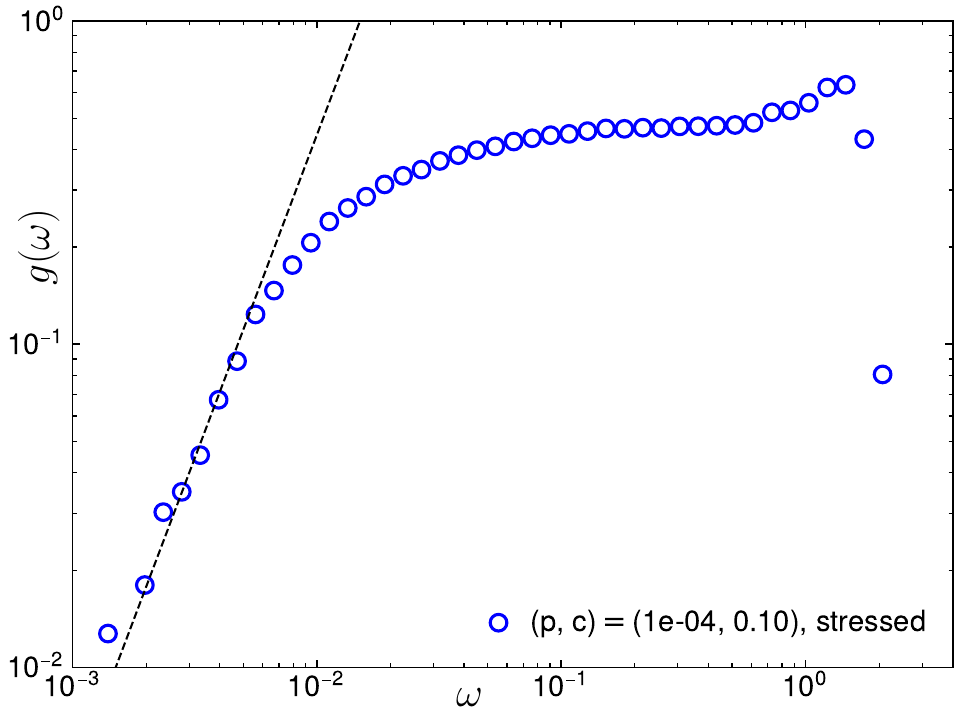}
\caption{The vibrational density of states of 2D packings obtained through the PC protocol.
The dashed line indicates $g(\omega) \propto \omega^2$.}
\label{fig:2Dnon-Debye}
\end{figure}

Additionally, we highlight that the non-Debye scaling has also been observed for the first time in 2D through the PC protocol.
In Fig.~\ref{fig:2Dnon-Debye}, we plot the VDOS of 2D packings generated by the PC protocol.
At low frequencies, the non-Debye scaling $g(\omega) \propto \omega^2$ is observed.
This is a novel result that can only be achieved by the introduction of pinning, as the past studies could only observe the non-Debye scaling in dimensions higher than three~\cite{Charbonneau_2016,Mizuno_2017,Shimada_2020}.
The presence of pinned particles can eliminate phonons at low frequencies that obscure the observation in two dimensions~\cite{Shiraishi2023pin2D}.
Moreover, the PC protocol introduced in this study can serve this role while maintaining the marginal stability of packings, and thus this observation is realized.
The result of Fig.~\ref{fig:2Dnon-Debye} reinforces the understanding that mean-field marginality has dimensional robustness and subsists in $d \geq 2$ dimensions.

\section{Quasi-localized modes}
\begin{figure}
\centering
\includegraphics[width=\linewidth]{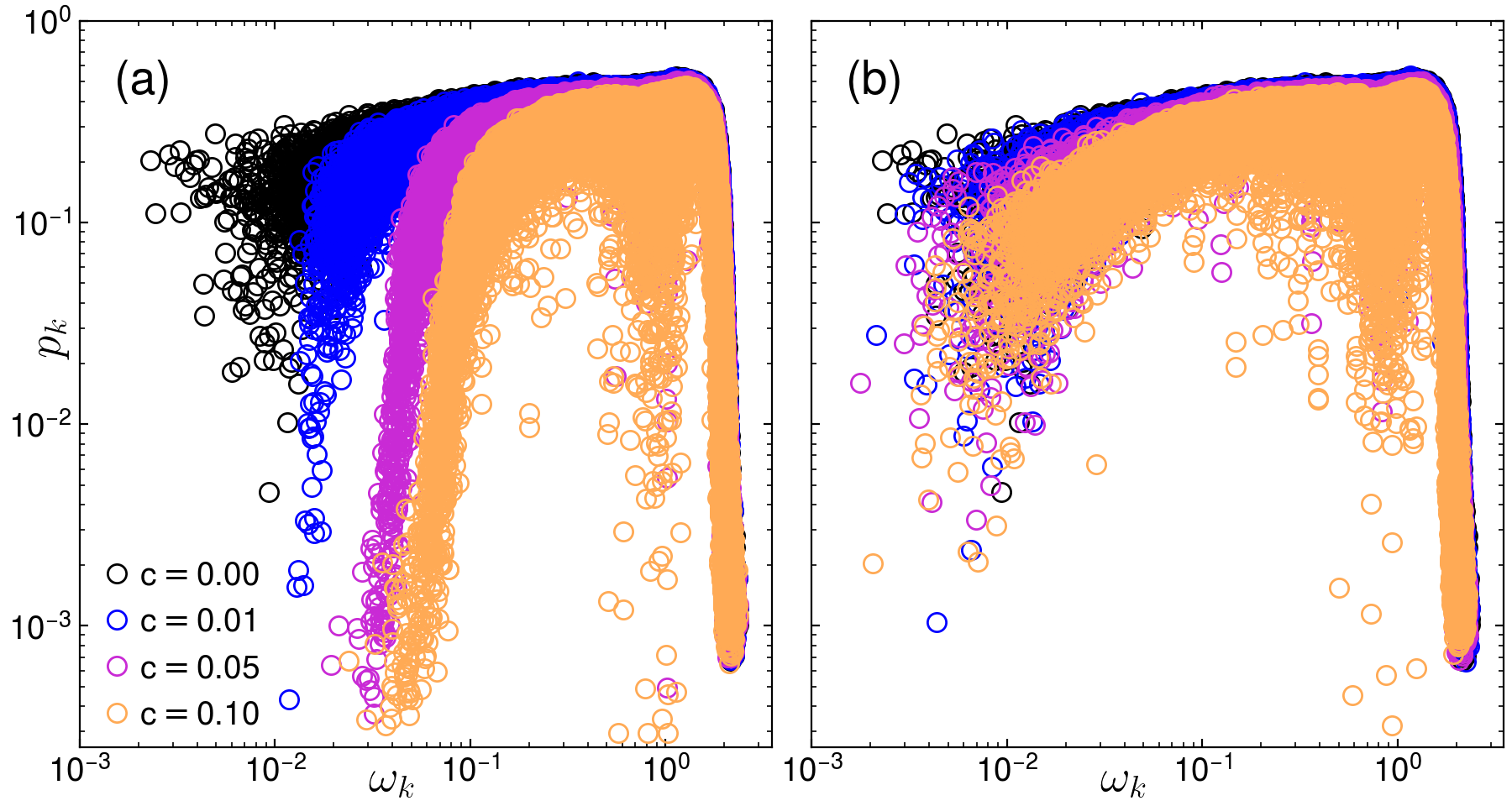}
\caption{The participation ratio of 3D packings ($N=4000$) at $p=\num{e-4}$ generated through (a) the CP protocol and (b) the PC protocol.}
\label{fig:PR3D}
\end{figure}

In the lower-frequency regime, quasi-localized modes emerge and coexist with phonons~\cite{Mizuno_2017}.
To assess the impact of pinning on them, we computed the participation ratio $p_k$ also for 3D packings.
For the unpinned system ($c=0.00$), quasi-localized modes manifest at low frequencies with $p_k \gtrsim \num{e-3}$ (Fig.~\ref{fig:PR3D}).
Their spatial structures are known to consist of a strongly vibrating core and a far-field elastic field~\cite{Shimada_Mizuno_Wyart_Ikeda_2018}.
In the CP packings, the introduction of pinned particles results in localized vibrations down to the single-particle level ($p_k \approx 1/\Nup$) in regions below the plateau (Fig.~\ref{fig:PR3D} (a)).
This behavior resembles the low-frequency vibrations observed in Lennard-Jones systems with pinning~\cite{Shiraishi2022Ideal,Shiraishi2023pin2D}, and they are considered to be bare localized modes free from hybridization with phonons.
Also, the lowest frequency edge of the modes increases as $c$ becomes larger in these packings.
In contrast, the degree of localization remains largely unchanged in the PC protocol when pinned particles are introduced, similar to the case of $c=0.00$ (Fig.~\ref{fig:PR3D} (b)).
In this case, we expect that phononic modes propagate by weaving their way through pins and hybridize with localized excitations.

Same as the non-Debye scaling, the presence of quasi-localized modes can be attributed to marginal stability, as they are absent in the unstressed system where the packing deviates from marginal stability~\cite{Mizuno_2017}.
Consistent with this understanding, our findings in Fig.~\ref{fig:PR3D} demonstrate that quasi-localized modes vanish in stable CP packings and are preserved in marginally stable PC packings.
As previously mentioned, localized modes observed in CP packings resemble those seen in Lennard-Jones systems, despite the use of a PC-like protocol in the latter~\cite{Shiraishi2022Ideal,Shiraishi2023pin2D}.
It would be interesting to investigate the reason for this apparent contradiction, presumably due to the absence of global compression in their protocol.

Thus far, we have presented understanding of the localization of low-frequency modes in the pinned system from the perspective of marginal stability: PC packings that preserve marginal stability exhibit quasi-localized modes, whereas CP packings that are strongly stable and are no longer at the marginality exhibit truly-localized modes.
These observations suggest that instability brought by contact force plays only minor role in the emergence of the truly-localized mode in CP packings.
In order to clarify this point, we also perform the unstressed analysis in the pinned system for the modes below $\omega^*$.

\begin{figure*}
\centering
\includegraphics[width=.85\linewidth]{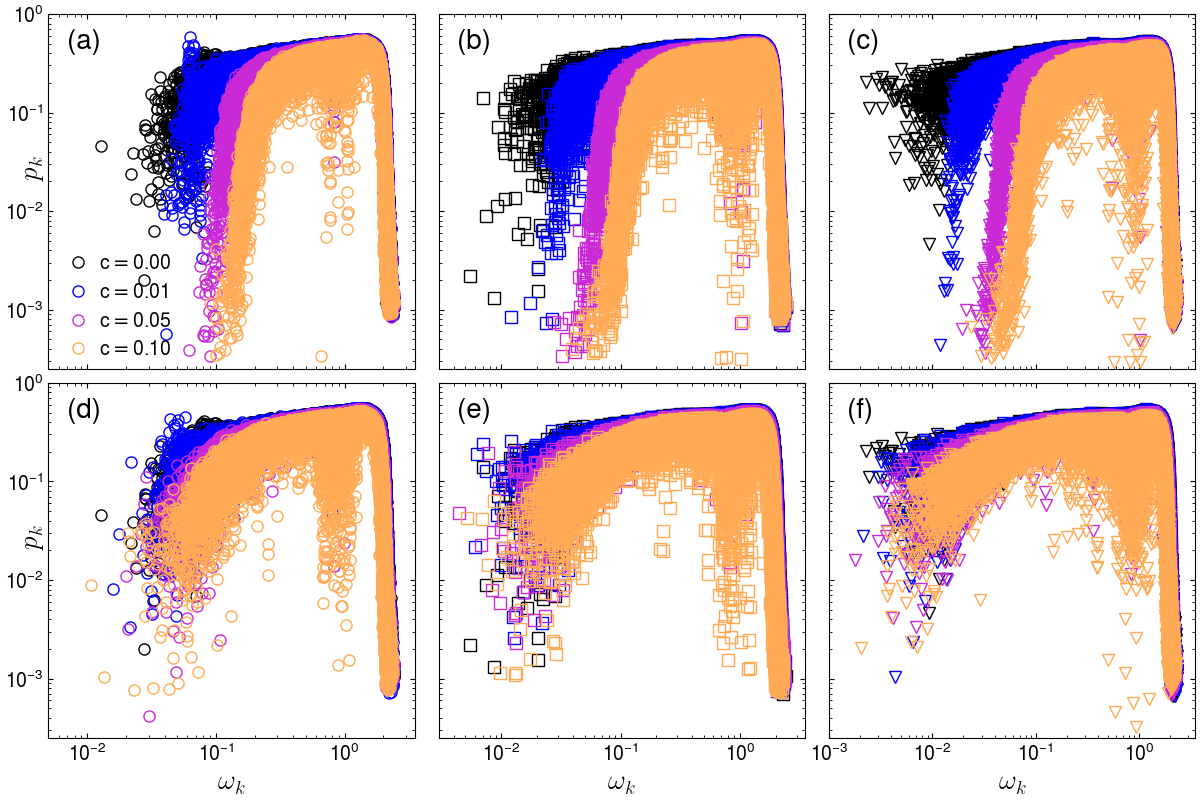}
\caption{Participation ratio of 3D packings in the stressed system.
Each panel presents the results for (a) $p=\num{e-2}$, (b) $p=\num{e-3}$, (c) $p=\num{e-4}$ of the CP protocol and (d) $p=\num{e-2}$, (e) $p=\num{e-3}$, (f) $p=\num{e-4}$ of the PC protocol.}
\label{fig:PR3Dstressed}
\end{figure*}

\begin{figure*}
\centering
\includegraphics[width=.85\linewidth]{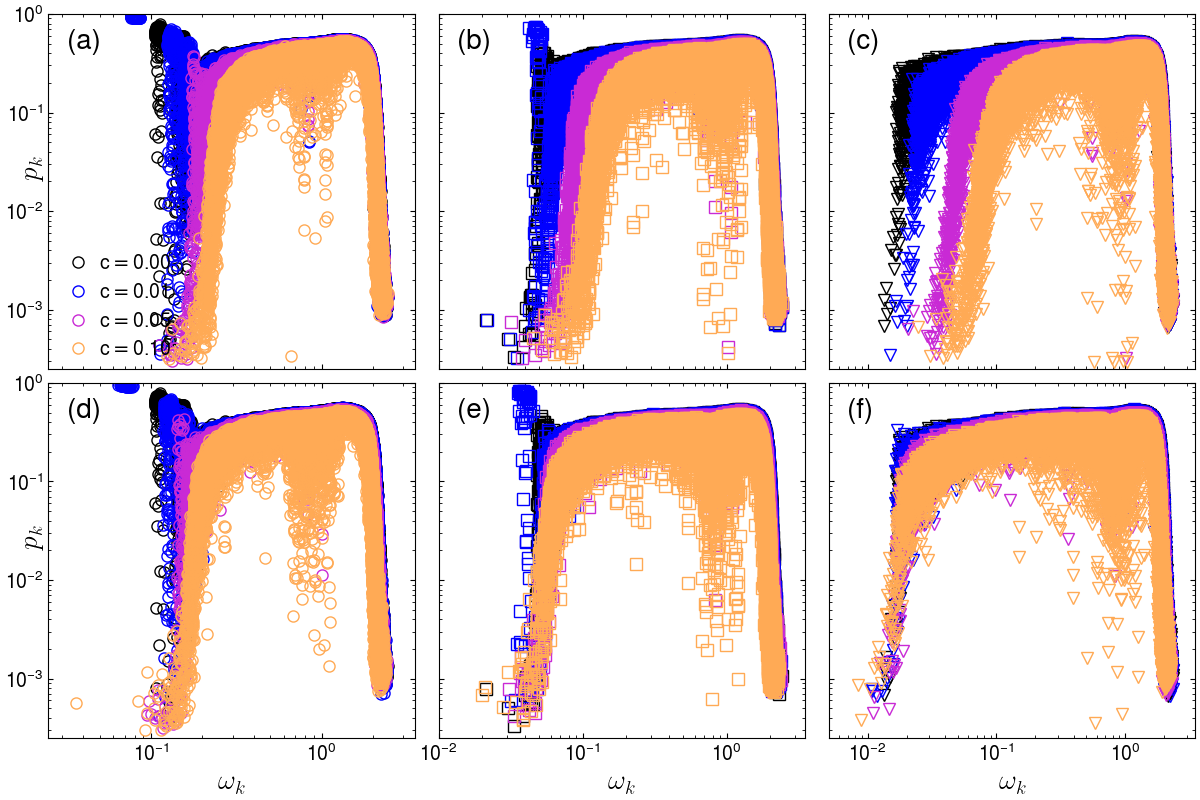}
\caption{Participation ratio of 3D packings in the unstressed system.
Each panel presents the results for (a) $p=\num{e-2}$, (b) $p=\num{e-3}$, (c) $p=\num{e-4}$ of the CP protocol and (d) $p=\num{e-2}$, (e) $p=\num{e-3}$, (f) $p=\num{e-4}$ of the PC protocol.}
\label{fig:PR3Dunstressed}
\end{figure*}

First, we present the participation ratio of the stressed system for multiple pressures and pinning fractions in Fig.~\ref{fig:PR3Dstressed}.
As we discussed with Fig.~\ref{fig:PR3D}, CP packings exhibit localized modes with $p_k \approx 1/\Nup$ at the low-frequency region for all pressures at large $c$ (Figs.~\ref{fig:PR3Dstressed} (a)--(c)).
On the other hand, for the PC protocol, $p_k$ of low-frequency modes do not deteriorate to $1/\Nup$: quasi-localized modes appear in these packings (Figs.~\ref{fig:PR3Dstressed} (d)--(f)).
Next, we present the participation ratio $p_k$ of the unstressed system in Fig.~\ref{fig:PR3Dunstressed}.
The localized vibrations with $p_k \approx 1/\Nup$ appear at the lowest-frequency region at large $c$ and, contrary to the stressed system, these strongly localized modes are observed not only in CP packings (Figs.~\ref{fig:PR3Dunstressed} (a)--(c)) but also in PC packings (Figs.~\ref{fig:PR3Dunstressed} (d)--(e)).
The stressed system of the CP configurations exhibits a similar behavior (Figs.~\ref{fig:PR3Dstressed} (a)--(c)), while the PC configurations show strongly localized modes only in the unstressed system.
This result indicates that the truly-localized vibrations that appear in low-frequency region of CP packings (Figs.~\ref{fig:PR3Dstressed} (a)--(c)) are not related to marginal stability because they are unchanging in the unstressed system, and their appearance in CP packings with large $\delta z$ are solely a result of the pinning operation~\cite{Shiraishi2022Ideal,Shiraishi2023pin2D}.

Finally, we study the scaling exponent $\beta$ of the VDOS $g(\omega) \propto \omega^\beta$ exhibited by truly-localized modes of CP packings.
To detect this exponent $\beta$, we prepared 50 packings and performed the vibrational analysis to each packing.
The method of using a large ensemble of configurations to detect the low-frequency exponent of $g(\omega)$ of glasses is widely used for both unpinned~\cite{Lerner_2016} and pinned systems~\cite{Angelani_2018,Shiraishi2022Ideal}.

\begin{figure}
\centering
\includegraphics[width=\linewidth]{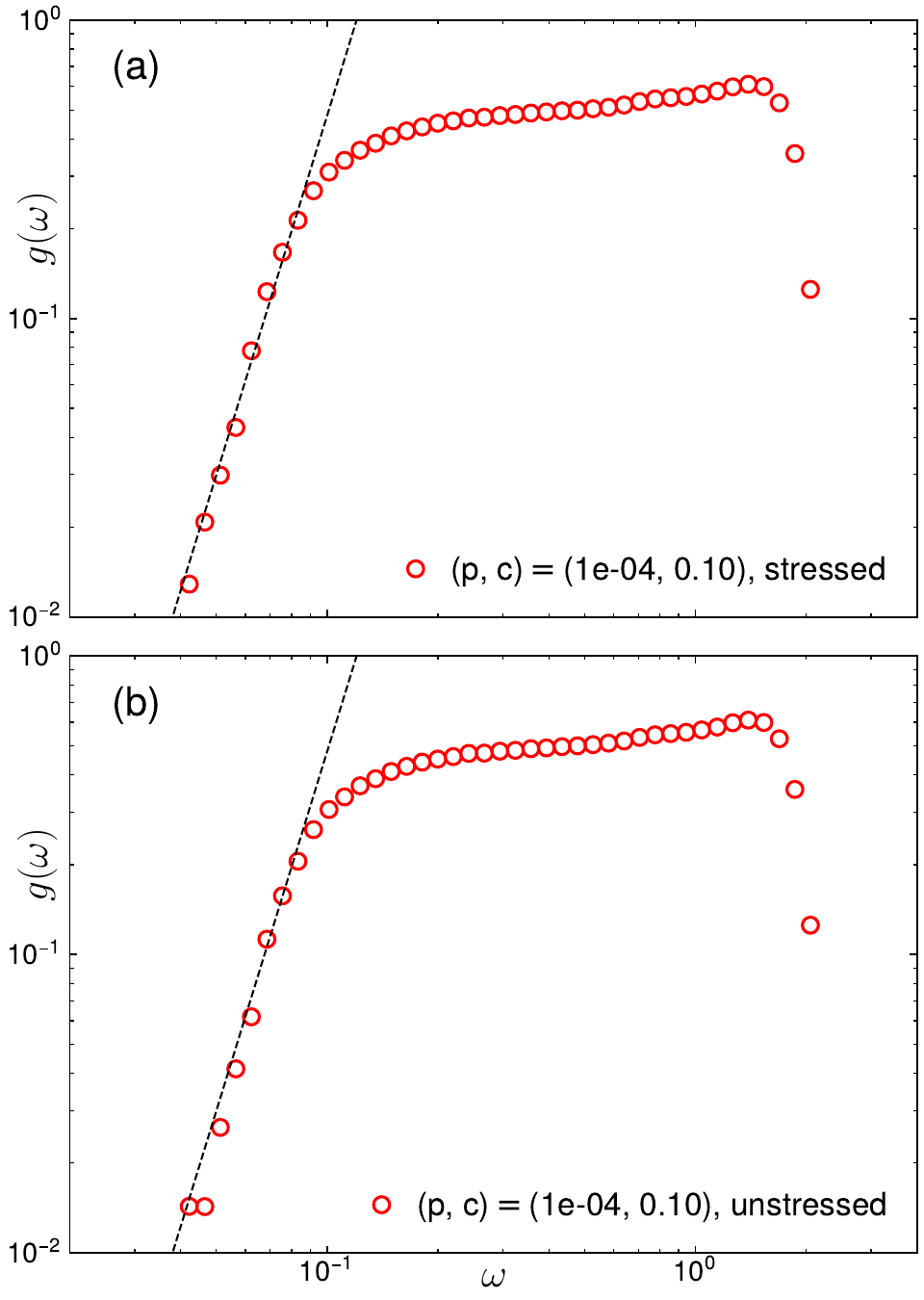}
\caption{The vibrational density of states of 3D packings obtained through the CP protocol.
The data from the (a) stressed and (b) unstressed analysis is used.
The dashed lines indicate $g(\omega) \propto \omega^4$.}
\label{fig:TrulyLocalizedVDOS}
\end{figure}

In Fig.~\ref{fig:TrulyLocalizedVDOS}, we show $g(\omega)$ of packings at $(p, c) = (\num{e-4}, 0.10)$ generated through the CP protocol.
$g(\omega)$ clearly shows the scaling of $g(\omega) \propto \omega^4$ at low frequencies, indicating that $\beta=4$ in this situation.
As discussed in Ref.~\cite{Shiraishi2023pin2D}, the random pinning method can be used to extract localized excitations that are inherent in glass configurations.
Figure \ref{fig:TrulyLocalizedVDOS} demonstrates that $\beta=4$ in 3D packings generated by the CP protocol for both (a) stressed and (b) unstressed systems and these values are consistent with the results of pinned Lennard-Jones~\cite{Shiraishi2022Ideal} and soft-spheres systems~\cite{Angelani_2018}.
From this coincidence, we conclude that excitations observed in Fig.~\ref{fig:TrulyLocalizedVDOS} are localized defects in glass configurations unrelated to the marginal stability.
The statistics of these excitations universally follow $g(\omega) \propto \omega^4$.

In previous studies, quasi-localized vibrations that constitute the contribution of $g(\omega) \propto \omega^4$ have been understood as a consequence of marginal stability:
modes that form the plateau in the unstressed system are destabilized by the effect of pre-stress (the term proportional to forces in the dynamical matrix) and appear as quasi-localized vibrations in lower frequencies in the stressed system~\cite{Wyart_PRE_2005,DeGiuli_2014,Mizuno_2017,Shimada_Mizuno_Wyart_Ikeda_2018}.
In fact, quasi-localized vibrations were observed in the marginally stable stressed system, but were absent in the unstressed system that is far from the marginally stable state~\cite{Mizuno_2017}.
This understanding naturally does not apply to the results of Fig.~\ref{fig:TrulyLocalizedVDOS}, because the localized modes with $g(\omega) \propto \omega^4$ emerge not only in the stressed system but also in the unstressed system in which the pre-stress terms are neglected.
Thus, the scaling $g(\omega) \propto \omega^4$ of CP packings in Fig.~\ref{fig:TrulyLocalizedVDOS} deviates from the past understanding~\cite{Wyart_PRE_2005,DeGiuli_2014,Mizuno_2017,Shimada_Mizuno_Wyart_Ikeda_2018}.
Future studies should be performed to understand the connection between the truly-localized excitations in CP packings and quasi-localized modes in ordinary/PC packings that are manifested by the marginal stability~\cite{Mizuno_2017}.

\section{Conclusion}
Our findings shed light on the influence of particle pinning on jammed solids in the context of marginal stability.
While the act of pinning particles freezes degrees of freedom and arithmetically alters their count, the behavior of $\omega^*$ regulated by isostaticity remains unaffected.
Our numerical simulations confirm that modifications occur to marginal stability, leading to consistent alterations in vibrations below $\omega^*$.
Recalling the stability phase diagram depicted in Fig.~\ref{fig:phase_diagram}, it highlights particle pinning as a novel method for manipulating marginal stability.
Thus far, the method employed to this end is the unstressed analysis, which drops pre-stressed terms in the dynamical matrix and has been implemented in simulations~\cite{Wyart_PRE_2005,DeGiuli_2014,Mizuno_2017}.
In the phase diagram, this method corresponds to stabilizing packings that lie in the instability line $\delta z=Cp^{1/2}$ along $p$-axis ($\delta z > Cp^{1/2} = 0$).
Our findings present a new possibility by demonstrating that pinning can exert control of stabilizing packings in $z$-axis as depicted in Fig.~\ref{fig:phase_diagram} ($\delta z \approx 2dc > Cp^{1/2}$), expanding the scope of manipulating the stability of jammed materials.
Furthermore, this new method has the advantage that the stability can be operationally decided thorough the fraction $c$ of pinned particles.
The capacity to exert deliberate control over marginal stability and low-frequency vibrations through the experimentally viable~\cite{Gokhale_2014,Hima_Nagamanasa_2015,Chen2023} method holds the promise of influencing the properties of amorphous solids, as these vibrations play a pivotal role in governing diverse properties such as transport~\cite{Xu_2009,Vitelli_2010,Mizuno_2018} or plastic rearrangements~\cite{Tanguy_2010,Manning_2011}.
Such control opens up avenues for tailoring mechanical and thermal behaviors of these materials.

\begin{acknowledgments}
We thank Hideyuki Mizuno for critically reading the manuscript and helpful discussions.
This work is supported by the Japan Society for the Promotion of Science (JSPS) through the Grant-in-Aid for JSPS Fellows (23KJ0368, Y.H.) and the Overseas Research Fellowships (K.S.).
\end{acknowledgments}

\end{document}